\documentclass[12pt]{article}
\pdfoutput=1
\usepackage{fancyvrb}
\usepackage{amsmath}
\usepackage{amssymb}
\usepackage{graphicx}
\usepackage[colorlinks,linkcolor=black,anchorcolor=black,citecolor=blue]{hyperref}
\usepackage[bottom]{footmisc}
\usepackage[top=1in, bottom=1in, left=1in, right=1in]{geometry}

\fvset{frame=none,framerule=0.1pt,framesep=6pt,numbers=left,xleftmargin=25pt,fontfamily=tt,fontsize=\small}
\newenvironment{hanging}
    {\begin{list}{}{\setlength\itemsep{0pt}%
 \setlength\topsep{0pt}%
 \setlength\leftmargin{25pt}%
 \setlength\itemindent{0pt}%
 \setlength\listparindent{\itemindent}}%
     \item[]}
    {\end{list}}

\newenvironment{cdbin}{\fvset{firstnumber=1}\Verbatim}{\endVerbatim}
\newenvironment{cdbcont}{\fvset{firstnumber=last}\Verbatim}{\endVerbatim}

\begin{document}

\title{\boldmath Computations of superstring amplitudes in pure spinor formalism via Cadabra}

\author{Ke-Sheng Sun$^a$\footnote{sunkesheng@amss.ac.cn and sunkesheng@126.com}
,\; Xiang-Mao Ding$^b$\footnote{xmding@amss.ac.cn},\; Fei Sun$^c$\footnote{sunfei@ctgu.edu.cn},\; Hai-Bin Zhang$^d$\footnote{hbzhang@hbu.edu.cn}\\
$^a$National Center for Mathematics and Interdisciplinary Sciences, AMSS,\\
CAS,Beijing 100190, China\\
$^b$Institute of Applied Mathematics, AMSS,\\
CAS,Beijing 100190, China\\
$^c$Department of Physics, China Three Gorges University,\\
Yichang 443002, China\\
$^d$Department of Physics, Hebei University,\\
Baoding 071002, China}
\date{}

\maketitle

\begin{abstract}
In this paper, we will illustrate how computer algebra system Cadabra is used in computing the supersymmetric amplitude in pure spinor formalism and provide the source code that computes the tree-level massless 5-gluon amplitude.

\end{abstract}

\eject
\pagestyle{plain}
\hrule
\tableofcontents
\vspace{4ex}
\hrule
\bigskip
\section{Introduction}
\label{sec:1}

The understanding of structure of superstring scattering amplitudes has gained great progresses since  discovery of the pure spinor formalism \cite{NB2000}, which has manifest spacetime supersymmetry and can be quantized covariantly. Besides of the usual superspace variables, a new bosonic spacetime spinor $\lambda^{\alpha}$ is introduced as the worldsheet ghosts and satisfies the pure spinor constraint
\begin{eqnarray}
\lambda^{\alpha} \gamma^{m}_{\alpha\beta}\lambda^{\beta}=0.
\end{eqnarray}
The origins of pure spinor are explained in \cite{NB2008,NBDM,YAYK,NB2015}. The equivalence of four-point tree-level, one-loop and two-loop amplitudes with the RNS results are proved in \cite{NBBV2000,CM2006,NBCM2006PRL}. The five- and six-point tree amplitudes have been explicitly computed in \cite{CM2010,MB2006,CM2011}.  For color-ordered n-point tree-level amplitudes, the recursive method is presented by the pure spinor superspace cohomology formula \cite{COSD2011D83}. Here we note that a first closed formula for the open superstring five-point amplitude using the RNS formalism is done in \cite{MBM2002,MB2005} and the n-point closed formula for tree level scattering of open superstrings using the RNS formalism is done in \cite{BM2014}.

The massless unintegrated vertex operator $V$ and integrated vertex operator $U$ in pure spinor formalism are described by the single-particle superfields $K\in\{A_{\alpha}$, $A^{m}$, $W^{\alpha}$, ${\cal F}_{mn}\}$. The CFT correlator in Eq.(\ref{CFTc}) can always be expressed in terms of BRST building blocks $T_{123...n}$ \cite{COSD2011D83,COS2013873}. In order to get the recursion relations for scattering amplitudes at tree-level, the BRST building blocks $T_{123...n}$ are grouped into the Berends-Giele currents $M_{12...n}$ \cite{BG1988}. The $n$-point SYM tree-level amplitude in pure spinor formalism is given by
\begin{eqnarray}
A^{\rm{SYM}}(1,2,...,n)=\langle E_{1...(n-1)}V^{n}\rangle,
\label{amp}
\end{eqnarray}
where the superfields $E_{1...(n-1)}$ is the sum of Berends-Giele currents $M_{12...n}$ and defined in Eq.(\ref{En}). In the computations of Eq.(\ref{amp}), the computation involving keeping the terms proportional to $\langle\lambda^{3}\theta^{5}\rangle$ and the contractions between terms with vector indices or Weyl spinor indices is tedious and impossible to be done by hand for higher-point or multiloop amplitudes. 

The complicated computations in superstring amplitudes can be carried out easily with the help of the computers. There are many packages aimed on algebraic manipulations with tensorial expressions, such as xAct\cite{xact}, Redberry\cite{redberry}, Maple Physics\cite{maple}, etc. In this paper, we will use a standalone package called Cadabra to deal with the calculations of superstring amplitudes. Cadabra is a computer algebra system aimed at, mainly but not restricted to, theoretical high energy physicists to deal  with tensorial mathematical expressions encountered in field theory \cite{KP1, KP2}. The input format and the output format are a subset of TeX which makes it user-friendly and the results readable. Cadabra knows about the concept of dummy indices therefor no special wildcard notation is needed. It features a substitution command which handles the anti-commuting objects and dummy indices. One can have precise control over the results by adding any arbitrary simplification step to a list of commands which is executed at every step.

In this paper, we will give an introduction on how the symbolic computer algebra system Cadabra is used in the calculations of superstring amplitudes with pure spinor formalism with explicit examples. In general, terms handled in the article include the definitions of the superfields $A_{\alpha}$, $A_{m}$, $W_{\alpha}$, ${\cal F}_{mn}$, the BRST Building blocks $T_{12...n}$, the methods of selecting terms proportional to $\langle\lambda^{3}\theta^{5}\rangle$, the realization of the contractions between terms with vector indices or Weyl spinor indices, the description of the pure spinor correlations in terms of Kronecker deltas and epsilon tensors, and so on. We also display the complete source code for the calculations of the tree-level massless 5-gluon amplitude with detailed interpretations at every step.

The article is organized as follows. In section \ref{sec:2}, we give a simple review on how the tree-level n-point superstring amplitude is computed in pure spinor formalism. In section \ref{sec:3}, we will give an introduction on most useful algorithms modules in Cadabra, and illustrate these modules with many basic calculations encountered in the amplitude computation. The complete source code for the tree-level massless 5-gluon amplitude is presented in section \ref{sec:4}. The conclusion is drawn in section \ref{sec:5}.

\section{Review of the pure spinor formalism}
\label{sec:2}
In this section, we will give a brief review of the pure spinor formalism for the superstring, focusing on the n-point tree-level scattering amplitudes calculation. 

The computation of n-point tree-level scattering amplitudes is the evaluation of the correlation function with three unintegrated vertices $V^{i}$ and $n - 3$ integrated vertices $U^{i}$ \cite{NB2000}
\begin{eqnarray}
\langle V^{1}(z_{1})V^{n-1}(z_{n-1})V^{n}(z_{n})U^{2}(z_{2})...U^{n-2}(z_{n-2})\rangle,
\label{CFTc}
\end{eqnarray}
where $V^{i}$ and $U^{i}$ have conformal weight zero and one, respectively. $V^{i}$ and $U^{i}$ are conformal fields on the worldsheet parametrized by a complex coordinate $z$, and described by the superfields $\{A_{\alpha}$, $A^{m}$, $W^{\alpha}$, ${\cal F}_{mn}\}$,
\begin{eqnarray}
V^{i}&=&\lambda^{\alpha}A^{i}_{\alpha},\label{Vi}\\
U^{i}&=&\partial{\theta^{\alpha}}A^{i}_{\alpha}+\Pi^{m}A^{i}_{m}+d_{\alpha}W^{\alpha}_{i}+\frac{1}{2}N^{mn}{\cal F}^{i}_{mn},\label{Ui}
\end{eqnarray}
where the vector indices are taken from Latin alphabet $a, b, c,...= 0, 1, ..., 9$, and the Weyl spinor indices are taken from Greek alphabet $\alpha, \beta,... = 1, 2, ...,16$. $\theta^{\alpha}$ is the right handed Majorana Weyl spinor, $\Pi^{m}$ is the supersymmetric momentum, $d_{\alpha}$ is the conjugate momentum to $\theta^{\alpha}$ and $N^{mn}$ is the Lorentz current. The unintegrated vertex $V^{i}$ is BRST closed, i.e., $QV^{i}$ = 0, where $Q=\lambda^{\alpha}D_{\alpha}$ is the BRST operator and satisfies $Q^{2}$ = 0, if the superfield $A_{\alpha}$α is on shell. The integrated vertex $U^{i}$ satifies $QU^{i}$ = $\partial V^{i}$. The supersymmetric derivative $D_{\alpha}$ is given by
\begin{eqnarray}
D_{\alpha}=\frac{\partial}{\partial\theta^{\alpha}}+\frac{1}{2}(\gamma^{m}\theta)_{\alpha}\partial_{m}.
\end{eqnarray}

\begin{figure}
\begin{center}
\includegraphics[width=0.8\columnwidth]{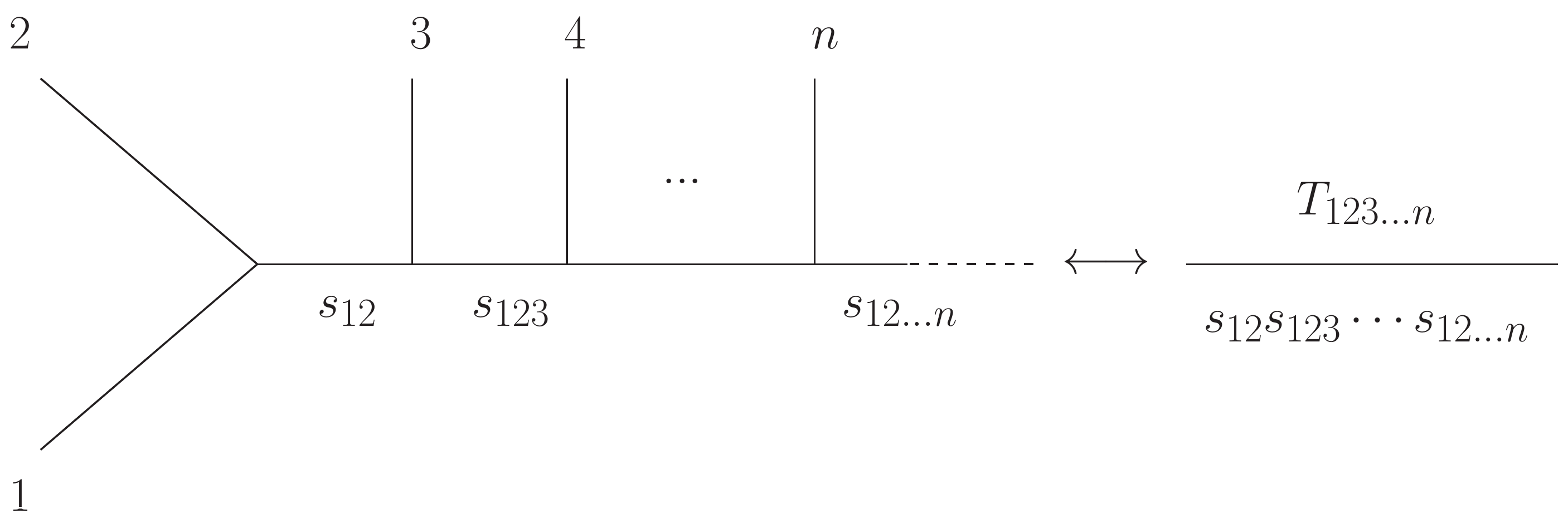}
\caption[]{Correspondence between cubic diagrams and the BRST building blocks $T_{12...n}$.}
\label{fig1}
\end{center}
\end{figure}

In the gauge $\theta^{\alpha}A_{\alpha}=0$, the $\theta$ expansions of ${\cal N}=1$ SYM superfields $A_{\alpha}$, $A_{m}$, $W_{\alpha}$, ${\cal F}_{mn}$ are given by \cite{GT2004,PD2006,HS1986}
\begin{eqnarray}
A_{\alpha}(x,\theta)&=&\frac{1}{2}a_{m}(\gamma^{m}\theta)_{\alpha}-\frac{1}{3}(\xi\gamma_{m}\theta)(\gamma^{m}\theta)_{\alpha}
-\frac{1}{32}F_{mn}(\gamma_{p}\theta)_{\alpha}(\theta\gamma^{mnp}\theta)+...,\label{Aa}\\
A_{m}(x,\theta)&=&a_{m}-(\xi\gamma_{m}\theta)-\frac{1}{8}(\theta\gamma_{m}\gamma^{pq}\theta)F_{pq}
+\frac{1}{12}(\theta\gamma_{m}\gamma^{pq}\theta)(\partial_{p}\xi\gamma_{q}\theta)+...,\label{Am}\\
W_{\alpha}(x,\theta)&=&\xi_{\alpha}-\frac{1}{4}(\gamma^{mn}\theta)_{\alpha}F_{mn}+\frac{1}{4}(\gamma^{mn}\theta)_{\alpha}
(\partial_{m}\xi\gamma_{n}\theta)+\frac{1}{48}(\gamma^{mn}\theta)_{\alpha}
(\theta\gamma_{n}\gamma^{pq}\theta)\partial_{m}F_{pq}+...\label{Wa},\\
{\cal F}_{mn}(x,\theta)&=&F_{mn}-2(\partial_{[m}\xi\gamma_{n}\theta)+\frac{1}{4}(\theta\gamma_{[m}\gamma^{pq}\theta)
\partial_{n]}F_{pq}+\frac{1}{6}\partial_{[m}(\theta\gamma_{n]}^{\;\; pq}\theta)(\xi\gamma_{q}\theta)\partial_{p}+...,\label{Fmn}
\end{eqnarray}
where $a_{m}$ = $e_{m}\;e^{ik\cdot x}$ and $\xi_{\alpha}$ = $\chi_{\alpha}\;e^{ik\cdot x}$ are the gluon polarization vector and gluino wave function, $F_{mn}$ = 2$\partial_{[m}a_{n]}$ is the linearized field strength. One can see that, for the superfields $A_{\alpha}$ and $W_{\alpha}$, the gluon polarization vector $a_{m}$ couples with odd powers of $\theta$ and the gluino wave function $\xi_{\alpha}$ with even powers of $\theta$. While the superfield $A_{m}$ and ${\cal F}_{mn}$ have the opposite feature. Although the $\theta$ expansions terminate at the finite order ${\cal O}(\theta^{16})$, ${\cal O}(\theta^{5})$ is sufficient for extracting superfield components $a_{m}$ and $\xi_{\alpha}$ from amplitudes. The superfields $A_{\alpha}$, $A_{m}$, $W_{\alpha}$, ${\cal F}_{mn}$ satisfy the equations of motion \cite{EW1986,NBictp}
\begin{eqnarray}
&&D_{\alpha}A_{\beta}+D_{\beta}A_{\alpha}=\gamma^{m}_{\alpha\beta}A_{m},\;\;\;D_{\alpha}A_{m}=(\gamma W)_{\alpha}+k_{m}A_{\alpha},\nonumber\\
&&D_{\alpha} {\cal F}_{mn}=2k_{[m}(\gamma_{n]}W)_{\alpha}\;\;,\;\;\;D_{\alpha} W^{\beta}=\frac{1}{4} (\gamma^{mn})_{\alpha}^{\beta} {\cal F}_{mn},\nonumber
\end{eqnarray}
which are on shell constraints for the construction of vertex operators in the BRST cohomology. The operators in Eq.(\ref{Vi}) and Eq.(\ref{Ui}) satisfy the following OPEs
\begin{align}
d_{\alpha}(z_{i}) V(z_{j})&\rightarrow\frac{D_{\alpha} V(z_{j})}{z_{ij}},&
\Pi^{m}(z_{i})V(z_{j})&\rightarrow -\frac{k^{m}V(z_{j})}{z_{ij}},\nonumber\\
d_{\alpha}(z_{i})\Pi^{m}(z_{j})&\rightarrow\frac{(\gamma^{m}\partial\theta)_{\alpha}}{z_{ij}},&
d_{\alpha}(z_{i})d_{\beta}(z_{j})&\rightarrow-\frac{\gamma^{m}_{\alpha\beta}\Pi_{m}}{z_{ij}},\nonumber\\
\Pi^{m}(z_{i})\Pi^{n}(z_{j})&\rightarrow-\frac{\eta^{mn}}{z_{ij}^{2}},&
d_{\alpha}(z_{i})\theta_{\beta}(z_{j})&\rightarrow-\frac{\delta^{\beta}_{\alpha}}{z_{ij}},\nonumber\\
d_{\alpha}(z_{i})\partial\theta_{\beta}(z_{j})&\rightarrow\frac{\delta^{\beta}_{\alpha}}{z_{ij}^{2}},&
N^{mn}(z_{i})\lambda^{\alpha}(z_{j})&\rightarrow-\frac{1}{2}\frac{(\lambda\gamma^{mn})^{\alpha}}{z_{ij}}.\nonumber\\
N^{mn}(z_{i})N_{pq}(z_{j})&\rightarrow\frac{4}{z_{ij}}N^{[m}_{[p}\delta^{n]}_{q]}-\frac{6}{z_{ij}^{2}}\delta^{n}_{[p}\delta^{m}_{q]},&\nonumber
\end{align}
where $z_{ij}$ equal $z_{i}$ - $z_{j}$ and $V(z_{j})$ denotes the superfields $A_{\alpha}$, $A_{m}$, $W_{\alpha}$, ${\cal F}_{mn}$. When computing tree-level n-point amplitudes, the OPEs can be used to define composite superfields $L_{2131...n1}$ \cite{COS2013873}
\begin{eqnarray}
&&\lim_{z_{2}\rightarrow z_{1}}V^{1}(z_{1})U^{2}(z_{2})\rightarrow \frac{L_{21}(z_{1})}{z_{21}},\;\;
\lim_{z_{3}\rightarrow z_{1}}L_{21}(z_{1})U^{3}(z_{3})\rightarrow \frac{L_{2131}(z_{1})}{z_{31}}.\nonumber\\
&&\lim_{z_{p}\rightarrow z_{1}}L_{2131(n-1)1}(z_{1})U^{n}(z_{n})\rightarrow \frac{L_{2131...n1}(z_{1})}{z_{n1}}.\nonumber
\end{eqnarray}
At $n=2$ and $n=3$, the OPE residues $L_{2131...p1}$ are given by
\begin{eqnarray}
L_{21}&=&-A^{1}_{m}(\lambda\gamma^{m}W^{2})-V^{1}(k^{1}\cdot A^{2}),\label{L12}\\
L_{2131}&=&-L_{21}(k^{1}+k^{2})\cdot A^{3}+(\lambda\gamma^{m}W^{3})[A^{1}_{m}k^{1}\cdot A^{2}+A^{1}_{n}{\cal F}^{2}_{mn}-(W^{1}\gamma_{m}W^{2})],
\end{eqnarray}
where the BRST exact terms have been discarded cause they make the OPE residues $L_{2131...p1}$ lack of symmetry under exchange of labels $1,2,3,...,n$. The BRST exact terms decouple and are canceled out in the final superspace expressions for the 5- and 6-point computations\cite{CM2010,CM2011}, and it is strongly suggest that this pattern has to persist at higher points. The removal of the BRST exact terms leads to the so-called BRST building blocks $T_{123...n}$ \cite{COS2013873,COSD2011D83},
\begin{eqnarray}
T_{123...n}\equiv L_{2131...n1}-corrections,\nonumber
\end{eqnarray}
which transform covariantly under the action of the BRST charge $Q$. At $n= 2$ and $n= 3$, the building blocks $T_{123...n}$ are given by
\begin{eqnarray}
T_{12}&=& \frac{1}{2}(L_{21}-L_{12})\equiv L_{[21]},\label{T12}\\
T_{123}
&=& \frac{1}{3}(\tilde{T}_{123}-\tilde{T}_{213})- \frac{1}{6}(\tilde{T}_{321}-\tilde{T}_{312}+\tilde{T}_{132}-\tilde{T}_{231}),
\end{eqnarray}
with
\begin{eqnarray}
\tilde{T}_{123}=L_{2131}+\frac{s_{12}}{2}[(A^{1}\cdot A^{3})V^{2}-(A^{2}\cdot A^{3})V^{1}]+\frac{s_{13}+s_{23}}{2}(A^{1}\cdot A^{2})V^{3},\nonumber
\end{eqnarray}
where $s_{ij}$=$\frac{1}{2}(k^{i}+k^{j})^{2}$ denote the standard Mandelstam variables. The diagram interpretation for $T_{12...n}$ in terms of tree subdiagrams with cubic vertices is shown in Fig.\ref{fig1}.
The cubic diagrams with branches is associated with a sum of several BRST building blocks $T_{12...n}$ in which the labels $1,2,3,...,n$ have been exchanged. At each rank the BRST building blocks $T_{123...n}$ obey one new symmetry in its labels while still inheriting all the lower-rank symmetries \cite{COS2013873}. For example,
\begin{eqnarray}
&&T_{12}+T_{21}=0,(rank\;2),\;T_{123}+T_{231}+T_{312}=0,(rank\;3),\nonumber\\
&&T_{123...n}+T_{213...n}=0,(rank\;n> 2),\;T_{123...n}+T_{231...n}+T_{312...n}=0, (rank\;n> 3).\nonumber
\end{eqnarray}

\begin{figure}
\begin{center}
\includegraphics[width=0.9\columnwidth]{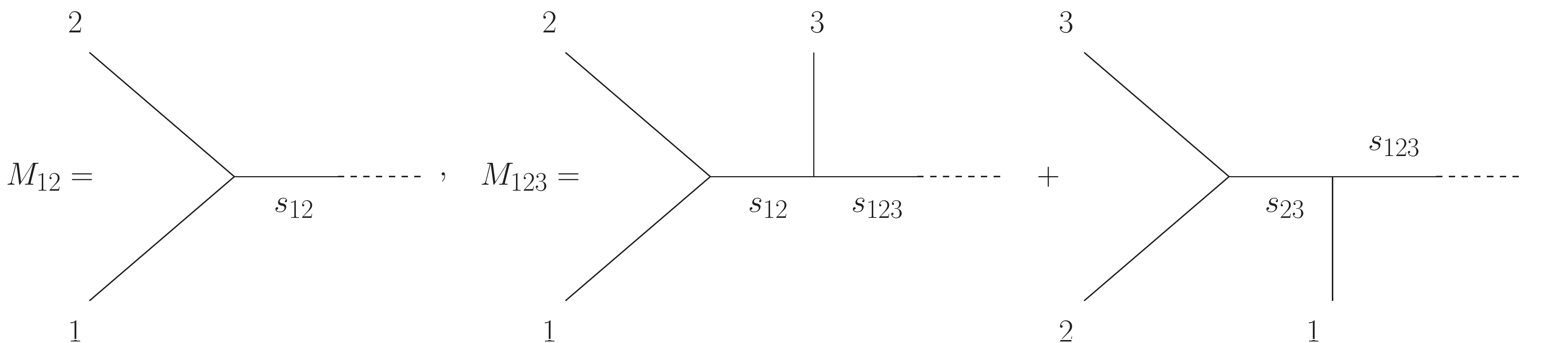}
\caption[]{The diagram interpretation of Berends-Giele currents $M_{12}$ and $M_{123}$.}
\label{fig2}
\end{center}
\end{figure}

The BRST building blocks $T_{123...n}$ are combined to get the so-called Berends-Giele currents $M_{123...n}$ \cite{BG1988}, which can be thought of as color-ordered SYM tree amplitudes with one leg off-shell. For the convenience of notation, $M_{i}$ is identified by $V_{i}$, $M_{i}\equiv V_{i}$. At $n= 2$ and $n= 3$, the Berends-Giele currents $M_{123...n}$ are given by
\begin{eqnarray}
M_{12}&=&\frac{T_{12}}{s_{12}},\nonumber\\
M_{123}&=&\frac{T_{123}}{s_{12}s_{123}}+\frac{T_{321}}{s_{23}s_{123}},\nonumber
\end{eqnarray}
which correspond to the three- and four-point amplitudes with one leg off-shell and the corresponding diagram interpretation is drawn in Fig.\ref{fig2}.
The Berends-Giele currents take several symmetry properties: the reflection identity, the cycle symmetry and the Kleiss-Kuijf relation \cite{KK1989}
\begin{eqnarray}
M_{12...n}&=&(-1)^{n-1}M_{n...21},\sum_{\sigma\in cyclic}M_{\sigma(1,2,...,n)}=0,\\
M_{\{\beta\},1,\{\alpha\}}&=&(-1)^{n_{\beta}}\sum_{\sigma\in OP(\{\alpha\},\{\beta^{T}\})}M_{1,\{\sigma\}}=0,
\label{}
\end{eqnarray}
where $OP(\{\alpha\},\{\beta^{T}\})$ denotes the set of all the permutations of $\{\alpha\}\bigcup \{\beta^{T}\}$ that maintain the order of the individual elements of both sets $\{\alpha\}$ and $\{\beta^{T}\}$, $n_{\beta}$ denotes the cardinality of set $\{\beta\}$, and $\{\beta^{T}\}$ denotes the set $\{\beta\}$ with reversed order of elements. Under the action of BRST charge $Q$, the Berends-Giele currents $M_{123...n}$ are translated into the bosonic superfield $E_{i_{1}...i_{n-1}}$ which is given by
\begin{eqnarray}
E_{i_{1}...i_{n-1}}=\sum^{n-2}_{i=1}M_{12...i}\;M_{(i+1)...(n-1)}.
\label{En}
\end{eqnarray}
The Berends-Giele currents give a compact description of the ten-dimensional SYM tree-level scattering amplitudes by \cite{COSD2011D83}
\begin{eqnarray}
A^{\rm{SYM}}(1,2,...,n)=\langle E_{i_{1}...i_{n-1}}V^{n}\rangle=\sum^{n-2}_{i=1}\langle M_{12...i}M_{(i+1)...(n-1)}V^{n}\rangle,
\label{Amp}
\end{eqnarray}
\begin{figure}
\begin{center}
\includegraphics[width=0.6\columnwidth]{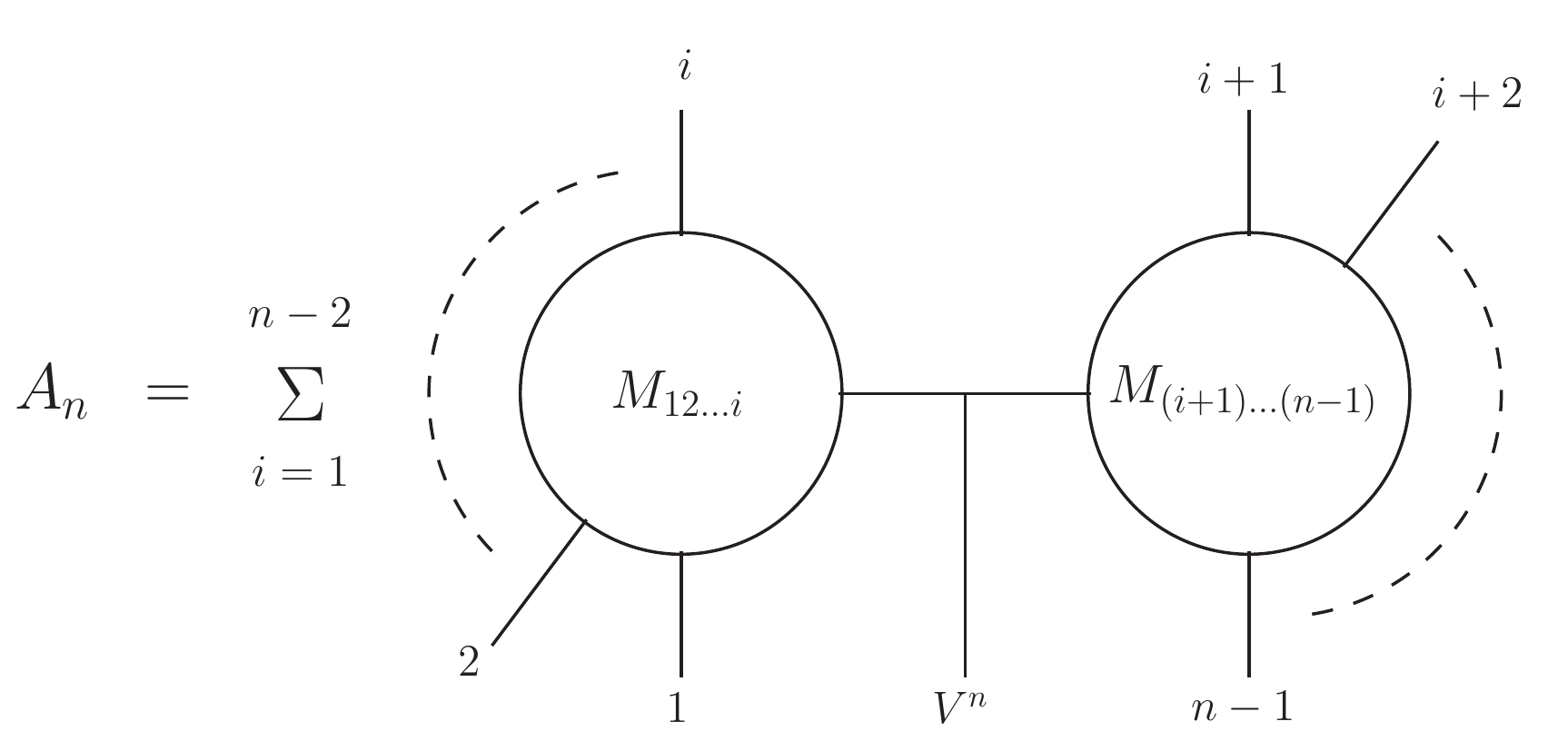}
\caption[]{The n-point SYM tree-level amplitude expressioned by Berends-Giele currents.}
\label{fig3}
\end{center}
\end{figure}
for which the diagram interpretation is shown in Fig.\ref{fig3}. 

After using the OPE’s to eliminate the conformal weight-one variables, the
integration of the zero-modes of $\lambda^{\alpha}$ and $\theta^{\alpha}$ is carried out by taking only the terms which contain three $\lambda$'s and five $\theta$'s in the correlator which are proportional to the zero-mode integration prescription \cite{NB2000}
\begin{eqnarray}
\langle (\lambda\gamma^{m}\theta)(\lambda\gamma^{n}\theta)(\lambda\gamma^{p}\theta)(\theta\gamma_{mnp}\theta)\rangle=2880.
\label{2880}
\end{eqnarray}
The choice of normalization factor 2880 is convenient in view of the factorization properties of pure spinor superspace kinematic factors and has been observed in \cite{NBBV2000} to imply tree-level normalizations compatible with RNS computations. Several pure spinor correlations proportional to Eq.(\ref{2880}), which will be used in examples in following sections, are given by in terms of Kronecker deltas and epsilon tensors \cite{NBCM2006}
\begin{eqnarray}
\langle(\lambda\gamma^{m}\theta)(\lambda\gamma^{s}\theta) (\lambda\gamma^{u}\theta) (\theta\gamma_{fgh}\theta)  \rangle&=& 24\;\delta^{smu}_{fgh}\label{PS1},\\
\langle(\lambda\gamma_{m}\theta)(\lambda\gamma_{s}\theta) (\lambda\gamma^{ptu}\theta) (\theta\gamma_{fgh}\theta)  \rangle&=&\frac{288}{7}\delta^{[p}_{[m}\delta_{s][f}\delta^{t}_{g}\delta^{u]}_{h]}\label{PS2},\\
\langle(\lambda\gamma_{m}\theta)(\lambda\gamma^{nrs}\theta) (\lambda\gamma^{ptu}\theta) (\theta\gamma_{fgh}\theta)  \rangle&=&\frac{12}{35}\epsilon^{fghmnprstu}+\frac{144}{7}\big[\delta^{[n}_{m}\delta^{r}_{[f}\delta^{s][p}\delta^{t}_{g}\delta^{u]}_{h]}
\nonumber\\ &&\hspace{-14em}-\delta^{[p}_{m}\delta^{t}_{[f}\delta^{u][n}\delta^{r}_{g}\delta^{s]}_{h]}\big]-\frac{72}{7}\big[\delta_{m[f}\delta^{v[p}\delta^{t}_{g}\delta^{u][n}\delta^{r}_{h]}\delta^{s]}_{v}-\delta_{m[f}\delta^{v[n}\delta^{r}_{g}\delta^{s][p}\delta^{t}_{h]}\delta^{u]}_{v}\big]
\label{PS3},
\end{eqnarray}
where $\delta^{msu}_{fgh}$ is the antisymmetrized combination of Kronecker deltas beginning with $\frac{1}{3!}\delta^{m}_{f}\delta^{s}_{g}\delta^{u}_{h}$. 

To illustrate how Cadabra is used in the SYM scattering amplitudes in pure spinor formalism, we will take the calculation of tree-level 5-gluon amplitude as an example in section \ref{sec:4}. At $n = 5$, the amplitude deduced from Eq.(\ref{Amp}) is 
\begin{eqnarray}
A_{5}&=&\langle E_{1234}V^{5}\rangle\nonumber\\
&=&\langle V^{1}M_{234}V^{5}\rangle+\langle M_{12}M_{34}V^{5}\rangle+\langle M_{123}V^{4}V^{5}\rangle\nonumber\\
&=&\frac{\langle T_{123}V^{4}V^{5}\rangle}{s_{12}s_{45}}-\frac{\langle T_{234}V^{1}V^{5}\rangle}{s_{23}s_{51}}
+\frac{\langle T_{12}T_{34}V^{5}\rangle}{s_{12}s_{34}}-\frac{\langle T_{231}V^{4}V^{5}\rangle}{s_{23}s_{45}}+\frac{\langle T_{342}V^{1}V^{5}\rangle}{s_{34}s_{51}}
\end{eqnarray}
and a manifestly cyclic-invariant form reads
\begin{eqnarray}
A_{5}=\langle M_{12}V^{3}M_{45}\rangle+cyclic(12345)=\frac{\langle T_{12}V^{3}T_{45}\rangle}{s_{12}s_{45}}+cyclic(12345).\label{tree5}
\end{eqnarray}
The final result is a product of Mandelstam invariants $s_{ij}$, momenta $k^{i}_{a}$ as well as the polarization vectors $e^{i}_{a}$ and/or the fermionic spinor wavefunctions $\chi^{i}_{\alpha}$. Using momentum conservation all 5-point kinematic invariants can be expressed in terms of \{$s_{12}$, $s_{23}$, $s_{34}$, $s_{45}$, $s_{15}$\} as
\begin{eqnarray}
&&s_{12}=s_{12},s_{23}=s_{23},s_{34}=s_{34},s_{45}=s_{45},s_{15}=s_{15},s_{13}=s_{45}-s_{12}-s_{23},\nonumber\\
&&s_{14}=s_{23}-s_{15}-s_{45},s_{24}=s_{15}-s_{23}-s_{34},s_{25}=s_{34}-s_{12}-s_{15},\nonumber\\
&&s_{35}=s_{12}-s_{45}-s_{34}.
\label{ss}
\end{eqnarray}

\section{Definition in Cadabra}
\label{sec:3}
The algorithms modules of Cadabra is built with three categories: \emph{Properties}, \emph{Algorithms} and \emph{Reserved node names}. \emph{Properties} assign properties to symbols. Symbols in Cadabra have no  priori meaning. If you write $\backslash \text{gamma}$, the program will not know that it is supposed to be a generator of a Clifford algebra. You will have to declare the properties of symbols by using the ``::\verb|GammaMatrix|'' as
\begin{cdbin}
\gamma{#}::GammaMatrix(metric=\delta).
\end{cdbin}
The \verb|GammaMatrix| property has turned the gamma
symbols into noncommuting objects, which will not change order when
sorting symbols in a product. \emph{Algorithms} can be made to act on existing expressions. This is done by:
\begin{cdbin}
@command(expression number or label){arg1}{arg2}...{argm}
\end{cdbin}
All of these commands act on the top of the argument subtree. You can make them
act subsequently on all nodes to which they apply by postfixing the name with an exclamation mark, as in
\begin{cdbin}
@command!(expression number or label){arg1}{arg2}...{argm}
\end{cdbin}
This will search the tree in pre-order style, applying the algorithm on every node to
which the algorithm applies.
If you want an algorithm to act until the expression no longer changes, use a double
exclamation mark ``\verb|!!|'' . To understand the difference between one ``\verb|!|' and two ``\verb|!!|', there are some examples. This code will exchang position of two indices 

\begin{cdbin}
A_{a b};
@substitute!(
\end{cdbin}
the output is
\begin{flalign}
\hspace{5ex}&A_{a b};&\nonumber\\
&A_{b a};&
\end{flalign}
and the following code will make Cadabra stuck in dead cycle,
\begin{cdbin}
A_{a b};
@substitute!!(
\end{cdbin}
Therefore a restart of Cadabra is needed. A list of all properties, algorithms and node names is available in the graphical interface through the Help menu. 

In Cadabra, the input lines always have to be terminated with either a  semi-colon ``\verb|;|'', a colon ``\verb|:|'' or a dot ``\verb|.|''.  The inputs and outputs are in exact one-to-one correspondence.  Those statements that finish with a semi-colon will display the outputs, and those statements that finish with a colon will not display the outputs.  It is also possible to save the output to a file, by appending a file name after the semi-colon, as
\begin{cdbin}
A_{a b}; "file name"
\end{cdbin}
A dot is always used at the end of statements while defining properties to symbols.  The line starting with a ``\verb|#|'' sign is considered to be a comment and will be ignored completely. The expressions in Cadabra can be given a label so you can refer to them again later, this is done by writing the label before the expression and a ``\verb|:=|'' in between, e.g.,
\begin{cdbin}
vertex:=\lambda_{\alpha} A_{\alpha};
@(vertex);
\end{cdbin}
the output is
\begin{flalign}
\hspace{5ex}
&\lambda_{\alpha} A_{\alpha};&\nonumber\\
&\lambda_{\alpha} A_{\alpha};&
\end{flalign}
\subsection{Node names}
\label{sec:3:1}

A small number of node names are reserved and always mean the same thing. the names of these reserved nodes can not be changed. In this subsection we will introduce two node names appeared in superstring amplitude computation.

\emph{$\backslash$indexbracket}

The node name denotes a group of objects with indices which are collectively written. The default properties are \verb|Distributable| and \verb|IndexInherit|, i.e., the object can be distributed over the terms with \verb|@distribute| and should inherit the indices of its child objects.
The node name would be used to describe the objects with Weyl spinor indices in $\theta$ expansions of the superfields $A_{\alpha}$, $A^{m}$, $W^{\alpha}$, ${\cal F}_{mn}$ in Eqs.(\ref{Aa}-\ref{Fmn}). The object with  one Weyl spinor index can be expressed as, e.g.,
\begin{cdbin}
\indexbracket{\gamma^{m}\theta}_{\alpha};
\end{cdbin}
or equivalent, using one parenthesis,
\begin{cdbin}
(\gamma^{m}\theta)_{\alpha};
\end{cdbin}
the output is 
\begin{flalign}
\hspace{5ex}
&(\gamma^{m}\theta)_{\alpha};&\nonumber
\end{flalign}
The object with suppressed Weyl spinor indices can be expressed as, e.g.,
\begin{cdbin}
\indexbracket{(\theta\gamma^{m}\theta)};
\end{cdbin}
the output is 
\begin{flalign}
\hspace{5ex}
&((\theta\gamma^{m}\theta));&\nonumber
\end{flalign}
This is used in the description of pure spinor correlations in Eqs.(\ref{PS1}-\ref{PS3}). 

It is noted worthwhile that the input format with double parentheses is not standard and the output is same as using one parenthesis, e.g.,
\begin{cdbin}
{m,n}::Indices(vector).
(\theta\Gamma^{m}\theta)+((\theta\Gamma^{m}\theta));
@collect_terms!(
\end{cdbin}
the output is
\begin{flalign}
\hspace{5ex}
&(\theta\Gamma^{m}\theta)+((\theta\Gamma^{m}\theta));&\nonumber\\
&2(\theta\Gamma^{m}\theta);&
\end{flalign}

\emph{$\backslash$cdot}

The node name denotes a dot product of vectors in which the contracted indices are suppressed. It displays as an infix dot.  In amplitude computation, we will use this to describe the contraction between the polarization vectors $e^{i}$ and the momenta $k^{j}$ in final expression, e.g.,
\begin{cdbin}
\cdot{e^{1}}{k^{2}};
\end{cdbin}
the output is
\begin{flalign}
\hspace{5ex}
&e^{1}\cdot k^{2};&\nonumber
\end{flalign}

\subsection{Properties}
\label{sec:3:2}
Properties of objects are declared when the objects first appear in the input. The property information is stored separately, and
further appearances of the objects will automatically share these properties. The objects can have more than one property attached to them. In this subsection we will introduce some properties appeared in superstring amplitude computation.

\emph{::Indices}

The property declares index names to be usable for dummy index purposes. It would be used to describe the vector indices and Weyl spinor indices in $\theta$ expansions of the superfields $A_{\alpha}$, $A^{m}$, $W^{\alpha}$, ${\cal F}_{mn}$ in Eqs.(\ref{Aa}-\ref{Fmn}). Typical usage are of the form
\begin{cdbin}
{a,b,c,d,e,a#}::Indices(vector).
{\alpha,\beta,\gamma,\delta,\alpha#}::Indices(spinor).
\end{cdbin}
This declares \{a,b,c,d,e,a\#\} and \{$\alpha$,$\beta$,$\gamma$,$\delta$,$\alpha\#$\} are indices in the dummy index set ``vector'' and ``spinor'' respectively, where a\# and  $\alpha\#$ mean the entire infinite set of objects a1, a2, a3, $\ldots$ and  $\alpha1$,  $\alpha2$, $\alpha3$, $\ldots$.

\emph{::Integer}

The property indicates that the object takes values in the integers. For the vector indices and Weyl spinor indices appeared in superfield $A_{\alpha}$, $A^{m}$, $W^{\alpha}$, ${\cal F}_{mn}$,  the optional range would be specified by 
\begin{cdbin}
{a,b,c,d,e,a#}::Integer(0..9).
{\alpha,\beta,\gamma,\delta,\alpha#}::Integer(1..16).
\end{cdbin}
This declares  \{a,b,c,d,e,a\#\} and \{$\alpha$,$\beta$,$\gamma$,$\delta$,$\alpha\#$\} take the integer value from 0 to 9 and 1 to 16 respectively. Note that the range can also be any length, e.g., 
\begin{cdbin}
{a,b,c,d,e,a#}::Integer(0..d-1).
\end{cdbin}
This declares the range is 0 to d-1.

\emph{::GammaMatrix}

The property indicates the object is a generalised generator of a Clifford algebra. Typical usage are of the form
\begin{cdbin}
\gamma{#}::GammaMatrix(metric=\delta).
\end{cdbin}
The statement declares the objects $\gamma^{\#}$ and $\gamma_{\#}$ are gamma matrices and the metric is $\delta$. The notation \{\#\} stands for the subscripts and superscripts with an arbitrary number of indices. With one vector index, the object satisfies
\begin{eqnarray}
\gamma^{m}\gamma^{n}+\gamma^{n}\gamma^{m}=2\delta^{m n}.\nonumber
\end{eqnarray}
The objects with two or more vector indices are defined as
\begin{eqnarray}
\gamma^{m_{1}\ldots m_{n}}=\gamma^{[m_{1}}\ldots \gamma^{m_{n}]},\nonumber
\end{eqnarray}
where the anti-symmetrisation includes a division by n!.

\emph{::AntiCommuting}

In general, all objects are commutative if not specified. The property declares the objects anticommuting. It works for objects with and without indices. e.g.,
\begin{cdbin}
{\indexbracket{\gamma^{a b c}\theta}_{\alpha},\indexbracket{\gamma^{a}
\theta}_{\alpha}}::AntiCommuting.
\indexbracket{\gamma^{b c d}\theta}_{\beta} \indexbracket{\gamma^{a}
\theta}_{\alpha};
@prodsort!(
\end{cdbin}
the output is
\begin{flalign}
\hspace{5ex}
&(\gamma^{bcd}\theta)_{\beta}(\gamma^{a}\theta)_{\alpha};&\nonumber\\
&-(\gamma^{a}\theta)_{\alpha}(\gamma^{bcd}\theta)_{\beta};&
\end{flalign}

\emph{::KroneckerDelta}

The property denotes a generalised Kronecker delta symbol. When the object carries two indices, it is the usual Kronecker delta.  When the object carries more than two indices, the object means
\begin{eqnarray}
\delta^{n_{1}n_{2}\ldots n_{k}}_{m_{1}m_{2}\ldots m_{k}}=\delta^{n_{1}}_{[m_{1}}\delta^{n_{2}}_{m_{2}}\ldots\delta^{n_{k}}_{m_{k}]}.\nonumber
\end{eqnarray}
To convert the Kronecker deltas with more than two indices to usual Kronecker delta, one would use \verb|@breakgendelta|. e.g., for the right hand side in Eq.(\ref{PS1}), 
\begin{cdbin}
\delta{#}::KroneckerDelta.
\frac{1}{120}\delta_{d}^{a}_{e}^{b}_{f}^{c};
@breakgendelta(
\end{cdbin}
the output is
\begin{flalign}
\hspace{5ex}
&\frac{1}{120} \delta_{def}^{abc};&\nonumber\\
&\frac{1}{720}\delta_{d}^{a}\delta_{e}^{b}\delta_{f}^{c}-\frac{1}{720}\delta_{d}^{a}\delta_{f}^{b}\delta_{e}^{c}-\frac{1}{720}\delta_{e}^{a}\delta_{d}^{b}\delta_{f}^{c}+\frac{1}{720}\delta_{e}^{a}\delta_{f}^{b}\delta_{d}^{c}+\frac{1}{720}\delta_{f}^{a}\delta_{d}^{b}\delta_{e}^{c}-\frac{1}{720}\delta_{f}^{a}\delta_{e}^{b}\delta_{d}^{c};&
\end{flalign}

\emph{::Weight}

The property attachs a labelled weight to an object, which can subsequently be used in the algorithms \verb|@keep_weight|. In amplitude computation, we use the following statement to label weights to objects, e.g.,
\begin{cdbin}
\lambda{#}::Weight(label=PS1,value=1).
\theta{#}::Weight(label=PS2,value=1).
{\xi,\xi{#}}::Weight(label=Fermion).
{Z^{#},F^{#}}::Weight(label=Boson).
\end{cdbin}
The first and second statements attach the pure spinor $\lambda^{\#}$, $\lambda_{\#}$ with weight label ``PS1'' and value one, and the worldsheet variable $\theta^{\#}$, $\theta_{\#}$ weight label ``PS2'' and value one. The third and forth statements attach the fermionic polarizations $\xi$, $\xi^{\#}$, $\xi_{\#}$ with weight label ``Fermion'', and the bosonic polarizations $Z^{\#}$, $Z_{\#}$, the field strength $F^{\#}$, $F_{\#}$ with weight label ``Boson'', where the value is not shown (the default set is one).
See the example after the algorithm \verb|@keep_weight| in section \ref{sec:3:3}. 

\emph{::SortOrder.}

The property determines the preferred order of objects when a \verb|@prodsort| command is used. One application is to keep the order of the pure spinor $\lambda$, the $\gamma$ matrices and the worldsheet variable $\theta$ in an indexbracket. e.g., with  associating the property to symbols list \{$\lambda$, $\gamma$, $\theta$\},
\begin{cdbin}
{\lambda,\gamma{#},\theta}::SortOrder.
\indexbracket{(\lambda\gamma^{a}\theta)};
@prodsort!(
\end{cdbin}
the output is
\begin{flalign}
\hspace{5ex}
&((\lambda\gamma^{m}\theta));&\nonumber\\
&@prodsort:\;not\;applicable.&\nonumber\\
&((\lambda\gamma^{m}\theta));&
\end{flalign}
while, without the first property statement, the output is
\begin{flalign}
\hspace{5ex}
&((\lambda\gamma^{m}\theta));&\nonumber\\
&((\gamma^{m}\lambda\theta));&
\end{flalign}

\subsection{Algorithms}
\label{sec:3:3}
The built-in algorithms can be made on tensorial expression for the manipulation of symmetrization, anti-symmetrization, contraction, renaming dummy indices, collecting identical terms, doing gamma matrix algebra in any dimension and so on. In this subsection we will introduce some algorithms appeared in superstring amplitude computation.

\emph{@asym}

The algorithm anti-symmetrises a product or tensor in the indicated objects. This works both with normal objects  as well as with indices. It is used in the description of pure spinor correlation in Eq.(\ref{PS2}) and Eq.(\ref{PS3}). e.g.,
\begin{cdbin}
\delta_{d}^{a}\delta_{e}^{b}\delta_{f}^{c};
@asym!(
\end{cdbin}
and the output is
\begin{flalign}
\hspace{5ex}
&\delta^{a}_{d}\delta^{b}_{e}\delta^{c}_{f};&\nonumber\\
&\frac{1}{6}\delta_{d}^{a}\delta_{e}^{b}\delta_{f}^{c}-\frac{1}{6}\delta_{d}^{a}\delta_{f}^{b}\delta_{e}^{c}-\frac{1}{6}\delta_{e}^{a}\delta_{d}^{b}\delta_{f}^{c}+\frac{1}{6}\delta_{e}^{a}\delta_{f}^{b}\delta_{d}^{c}+\frac{1}{6}\delta_{f}^{a}\delta_{d}^{b}\delta_{e}^{c}-\frac{1}{6}\delta_{f}^{a}\delta_{e}^{b}\delta_{d}^{c};&
\end{flalign}
There is also a algorithm ``\verb|@sym|'' to handle the symmetrisation which is available through the Help menu.

\emph{@distribute}

The algorithm translates a product of sums to a sum of products, and works on the objects which carry the ``\verb|::Distributable|'' property. The algorithm is widely used to deal with normal objects as well as with indices, and we do not give examples in the paper.

\emph{@combine}

The algorithm combines two objects with consecutive contracted indices into one object with an indexbracket. The algorithm is used to contract the pure spinor $\lambda$, gamma matrices $\gamma$ and worldsheet variable $\theta$ into pure spinor correlations. e.g.,
\begin{cdbin}
\lambda_{\alpha} (\gamma^{m}\theta)_{\alpha} (\lambda\gamma^{n})_{\beta}
(\gamma^{r s}\theta)_{\beta} \theta_{\gamma} (\gamma^{p t u}\theta)_{\gamma};
@combine!(
\end{cdbin}
the output is
\begin{flalign}
\hspace{5ex}
&\lambda_{\alpha}(\gamma^{m}\theta)_{\alpha}(\lambda\gamma^{n})_{\beta}(\gamma^{rs}\theta)_{\beta}\theta_{\gamma}(\gamma^{ptu}\theta)_{\gamma};&\nonumber\\
&((\lambda\gamma^{m}\theta))((\lambda\gamma^{n}\gamma^{rs}\theta))((\theta\gamma^{ptu}\theta));&
\end{flalign}

\emph{@eliminate\_kr}

The algorithm eliminates Kronecker delta symbols by performing index contractions and replaces contracted Kronecker delta symbols with the declared range over which the index runs. e.g.,
\begin{cdbin}
\delta{#}::KroneckerDelta.
k^{3}_{m}e^{1}_{r}e^{2}_{s}e^{3}_{n} \delta_{n r}\delta_{m s};
@eliminate_kr!(
\end{cdbin}
the output is
\begin{flalign}
\hspace{5ex}
&k^{3}_{m}e^{1}_{r}e^{2}_{s}e^{3}_{n} \delta_{n r}\delta_{m s};&\nonumber\\
&k^{3}_{s}e^{1}_{n}e^{2}_{s}e^{3}_{n};&
\end{flalign}

Note that the algorithm \verb|@eliminate_kr| can eliminate usual Kronecker deltas, i.e., with two indices, for Kronecker deltas with more than two indices, one should use algorithm \verb|@breakgendelta|.

\emph{@breakgendelta}

The algorithm converts generalised deltas to products of usual Kronecker deltas. See the examples after the property ``KroneckerDelta'' in section \ref{sec:3:2}.

\emph{@keep\_weight}

The algorithm keeps only those terms for which a product has the indicated weight. Weights are
declared by making use of the ``\verb|::Weight|'' property of symbols. E.g., considering the case in which we want to keep all terms with $\langle\lambda^{3}\theta^{5}\rangle$, this is done by
\begin{cdbin}
\lambda{#}::Weight(label=PS1).
\theta{#}::Weight(label=PS2).
\indexbracket{#}::WeightInherit(label=all,type=Multiplicative).
\indexbracket{(\lambda\gamma^{a}\theta)}\indexbracket{(\lambda\gamma^{b}
\theta)}\indexbracket{(\theta\gamma^{c d e}\theta)}\indexbracket{(\theta
\gamma^{f g h}\theta)}+\indexbracket{(\lambda\gamma^{a}\theta)}
\indexbracket{(\lambda\gamma^{b}\theta)}\indexbracket{(\lambda\gamma^{c d e}
\theta)}\indexbracket{(\theta\gamma^{f g h}\theta)}+F_{i j}e_{k}
\indexbracket{(\lambda\gamma^{a}\theta)}\indexbracket{(\lambda\gamma^{b}
\theta)}\indexbracket{(\lambda\gamma^{c d e}\theta)}\indexbracket{(\theta
\gamma^{f g h}\theta)}\indexbracket{(\theta\gamma^{i j k}\theta)};
@keep_weight!(
@keep_weight!(
\end{cdbin}
the output is
\begin{flalign}
\hspace{5ex}
&((\lambda\gamma^{a}\theta))((\lambda\gamma^{b}\theta))((\theta\gamma^{cde}\theta))((\theta\gamma^{fgh}\theta))+((\lambda\gamma^{a}\theta))((\lambda\gamma^{b}\theta))((\lambda\gamma^{cde}\theta))((\theta\gamma^{fgh}\theta))&\nonumber\\
&+F_{ij}e_{k}((\lambda\gamma^{a}\theta))((\lambda\gamma^{b}\theta))((\lambda\gamma^{def}\theta))((\lambda\gamma^{fgh}\theta))((\lambda\gamma^{ijk}\theta));&\nonumber\\
&((\lambda\gamma^{a}\theta))((\lambda\gamma^{b}\theta))((\lambda\gamma^{cde}\theta))((\theta\gamma^{fgh}\theta))+F_{ij}e_{k}((\lambda\gamma^{a}\theta))((\lambda\gamma^{b}\theta))((\lambda\gamma^{def}\theta))((\lambda\gamma^{fgh}\theta))&\nonumber\\
&((\lambda\gamma^{ijk}\theta));&\nonumber\\
&((\lambda\gamma^{a}\theta))((\lambda\gamma^{b}\theta))((\lambda\gamma^{cde}\theta))((\theta\gamma^{fgh}\theta));&
\end{flalign}
The statement in third row declares that the object inherits all weights of its child nodes.

\emph{@reduce\_gendelta}

The algorithm converts generalised delta symbols which take contracted indices to deltas with
fewer indices, according to the formula
\begin{eqnarray}
n!\delta^{a_{1}\ldots a_{n}}_{b_{1}\ldots b_{n}}\delta^{b_{1}\ldots b_{m}}_{a_{1}\ldots a_{m}}=\big[\prod_{i=1}^{m}(d-n+i)\big](n-m)!\delta^{a_{m+1}\ldots a_{n}}_{b_{m+1}\ldots b_{n}}.
\end{eqnarray}
e.g.,
\begin{cdbin}
{a,b,c,d,e}::Integer(0..9).
\delta{#}::KroneckerDelta.
\frac{1}{120} \delta_{c}^{a}_{d}^{b}_{e}^{c};
@reduce_gendelta(
\end{cdbin}
the output is
\begin{flalign}
\hspace{5ex}
&\frac{1}{120} \delta_{cde}^{abc};&\nonumber\\
&\frac{1}{45} \delta_{de}^{ab};&
\end{flalign}

\emph{@substitute}

The algorithm replaces objects with something else. The algorithm can do very complicated things. The first example is to deal with the contractions between momenta $k^{i}_{a}$ and the polarization vectors $e^{j}_{a}$,
\begin{cdbin}
e^{1}_{a}e^{2}_{b}e^{3}_{b}k^{4}_{a}k^{5}_{c}k^{6}_{c};
@substitute!!(
e^{?}_{a}e^{??}_{a}->\cdot{e^{?}}{e^{??}},
k^{?}_{a}k^{??}_{a}->\cdot{k^{?}}{k^{??}});
\end{cdbin}
the output is
\begin{flalign}
\hspace{5ex}
&e^{1}_{a}e^{2}_{b}e^{3}_{b}k^{4}_{a}k^{5}_{c}k^{6}_{c};&\nonumber\\
&e^{1}\cdot k^{4} e^{2}\cdot e^{3} k^{5}\cdot k^{6};&
\end{flalign}
The single question mark and double question mark mean the two objects at the right-hand of the arrow may have the same superscript or not. If replacing the single question mark with the double question mark (or replacing the double question mark with  the single question mark), the algorithm only works with two objects that share the same superscript. The second example is to deal with long substitution rule. The substitution rule can be written into a labelled expression and called when necessary. Taking the unintegrated massless vertex operators $V^{i}$ for example, 
\begin{cdbin}
{\alpha,\beta,\gamma}::Indices.
Vi:=V^{?}=\lambda_{\alpha}A^{?}_{\alpha}:
V^{1} V^{2} V^{3};
@substitute!(
\end{cdbin}
the output is
\begin{flalign}
\hspace{5ex}
&V^{1} V^{2} V^{3};&\nonumber\\
&\lambda_{\alpha}A^{1}_{\alpha}\lambda_{\beta}A^{2}_{\beta}
\lambda_{\gamma}A^{3}_{\gamma};&
\end{flalign}
\section{Code for 5-gluon amplitude}
\label{sec:4}

In this section, we will illustrate the application of Cadabra in computation of tree-level 5-gluon superstring amplitude. The source code of the program is freely available and successfully compiled on multiple platforms including Linux and MacOSX. The codes in this paper are written with Cadabra version 1.42 on Ubuntu system, and one can install this version by enabling the PPA. To get the output one need to copy the code into Cadabra and press the shift+enter key.
\begin{cdbin}
::PostDefaultRules( @@collect_terms!(
\end{cdbin}
The statement in 1st row has set the default rules that the algorithm \verb|@collect_terms| should be applied after every new input has been processed and active nodes have been executed.
\begin{cdbcont}
{\alpha,\beta,\gamma,\delta,\alpha#,\beta#,\gamma#}::Indices(spinor).
{\alpha,\beta,\gamma,\delta,\alpha#,\beta#,\gamma#}::Integer(1..16).
{a,b,c,d,e,f,g,h,l,m,n,p,q,r,s,t,u,v,a#}::Indices(vector).
{a,b,c,d,e,f,g,h,l,m,n,p,q,r,s,t,u,v,a#}::Integer(0..9).
\Gamma{#}::GammaMatrix(metric=\delta).
\delta{#}::KroneckerDelta.
\partial{#}::PartialDerivative.
\epsilon{#}::EpsilonTensor(delta=\delta).
r#::Symbol.
\theta{#}::ImplicitIndex.
\lambda{#}::Weight(label=PS1).
\theta{#}::Weight(label=PS2).
{Z^{#},F^{#}}::Weight(label=Boson).
\indexbracket{#}::WeightInherit(label=all,type=Multiplicative).
\partial{#}::WeightInherit(label=all,type=Multiplicative).
{\lambda,\Gamma{#},\theta}::SortOrder.
{\lambda_{\alpha},(\Gamma^{a}\theta)_{\alpha}}::SortOrder.
{O_{1}^{a},O_{1}^{a b c},O_{2}^{a b c}}::SortOrder.
{O_{1}^{a},O_{1}^{a b c}}::AntiCommuting.
{(\Gamma^{a}\theta)_{\alpha},(\Gamma^{a b}\theta)_{\alpha},
(\Gamma^{a b c}\theta)_{\alpha},\theta{#},
\indexbracket{(\lambda\Gamma^{a}\theta)},
\indexbracket{(\lambda\Gamma^{a b c}\theta)}}::AntiCommuting.
\end{cdbcont}
The 2nd row to the 5th row declare Latin alphabet and Greek alphabet are integers and take the properties of vector indices and Weyl spinor indices. The 6th row defines the symbols $\Gamma$ with indices are $16\times 16$ Pauli matrices in $D=10$. The 7th row defines the symbols $\delta$ with indices are generalised Kronecker deltas. The 8th row defines the symbols $\partial$ with indices are partial derivatives. The 9th defines the symbols $\epsilon$ with indices are fully anti-symmetric tensors. The 10th row states $r1$, $r2$, $...$, are symbols. The 11th row states the symbols $\theta$ carries implicit indices, although we know the indices are Weyl spinor indices, this is used to contract with symbols like $(\lambda\gamma^{m})_{\alpha}$ by the algorithm \verb|@combine|. The 12th row to 14th row associate the symbols $\lambda$, $\theta$, $Z$, $F$ with labelled weights. The 15th and 16th row declare that the  indexbrackets and the symbols $\partial$ have the same weights of its child nodes. The 17th row and 18th row set the order of the symbols $\lambda$, $\gamma$ and $\theta$ in indexbrackets. For convenience, the following shorthands are used
\begin{eqnarray}
(\lambda \Gamma_{a} \theta) or (\lambda \Gamma^{a} \theta)&\rightarrow& O_{1}^{a}\label{O1}\\
(\lambda \Gamma_{a b c} \theta) or (\lambda \Gamma^{a b c} \theta)&\rightarrow& O_{1}^{a b c}\label{O2}\\
(\theta \Gamma_{a b c} \theta)  or (\theta \Gamma^{a b c} \theta)&\rightarrow& O_{2}^{a b c}\label{O3}
\end{eqnarray}
which are anticommuting with each other and declared in 19th row and 20th row. Other anticommuting symbols are declared in the 21st row to the 24th row.
\begin{cdbcont}
Aalpha:=A^{?}_{\alpha}=\frac{1}{2} Z^{?}_{m} (\Gamma^{m}\theta)_{\alpha}
-\frac{1}{32} F^{?}_{m n} (\Gamma^{p}\theta)_{\alpha} 
\indexbracket{(\theta \Gamma^{m n p} \theta)}+\frac{1}{1152}
(\Gamma^{m} \theta)_{\alpha} \indexbracket{(\theta\Gamma^{m r s} \theta)}
\indexbracket{(\theta \Gamma^{s p q} \theta)} \partial_{r}{F^{?}_{p q}};
Am:=A^{?}_{m}=Z^{?}_{m}-\frac{1}{8}\indexbracket{(\theta\Gamma^{m p q}\theta)}
F^{?}_{p q}+\frac{1}{192}\indexbracket{(\theta\Gamma^{m r s}\theta)}
\indexbracket{(\theta\Gamma^{s p q}\theta)}\partial_{r}{F^{?}_{p q}};
Walpha:=W^{\alpha}_{?}=-\frac{1}{4} (\Gamma^{m n} \theta)_{\alpha}F^{?}_{m n}
+\frac{1}{48} (\Gamma^{m n}\theta)_{\alpha}
\indexbracket{(\theta\Gamma^{n p q}\theta)}\partial_{m}{F^{?}_{p q}};
\end{cdbcont}
The 25th row to 35th row give the input of the $\theta$ expansions of $A_{\alpha}$, $A_{m}$, $W_{\alpha}$ in Eqs.(\ref{Aa}-\ref{Wa}), where the purely fermionic part is not used in 5-gluon superstring amplitude calculation and omitted in the input. 
\begin{cdbcont}
Vk:=Vk^{?}_{a}->k^{?}_{a}V^{?}: 
Vi:=V^{?}=\lambda_{\alpha}A^{?}_{\alpha}:
L21:=L_{?? ?}->-A^{?}_{a} (\lambda\Gamma^{a})_{\alpha} W^{\alpha}_{??}
-Vk^{?}_{a}A^{??}_{a}:
T12:=T_{? ??}->1/2 L_{?? ?}-1/2 L_{? ??}:
\end{cdbcont}
The code has used the notation $Vk^{?}_{a}=k^{?}_{a}V^{?}$ and $V^{?}=\lambda_{\alpha}A^{?}_{\alpha}$ which is shown in the 36th row and the 37th row. The 38th row and the 39th row give the input of the superfield $L_{12}$ in Eq.(\ref{L12}) and the BRST building block $T_{12}$ in Eq.(\ref{T12}).
\begin{cdbcont}
V1t:=24 \delta_{a d}_{b e}_{c f}): @breakgendelta!(V1t):
V1:=V1^{a b c d e f}=@(V1t):
V2t:=\frac{288}{7} \delta_{c a}\delta_{d g}\delta_{e h} \delta_{b f}: 
@asym!(V2t){_{c},_{d},_{e}}: @asym!(V2t){_{f},_{g},_{h}}:
@asym!(V2t){_{a},_{b}}: V2:=V2^{a b c d e f g h}=@(V2t):
V3t1:=\frac{12}{35} \epsilon_{f g h m n p r s t u}: 
V3t2:=\frac{144}{7} (\delta_{n m}\delta_{r f}\delta_{t g}\delta_{s p}
\delta_{u h}-\delta_{p m}\delta_{t f}\delta_{r g}\delta_{u n}\delta_{s h}): 
@asym!(V3t2){_{p},_{t},_{u}}: @asym!(V3t2){_{f},_{g},_{h}}: 
@asym!(V3t2){_{n},_{r},_{s}}: V3t3:=\frac{72}{7} (\delta_{m f}\delta_{v p}
\delta_{t g}\delta_{u n}\delta_{r h}\delta_{s v}-\delta_{m f}\delta_{v n}
\delta_{r g}\delta_{s p}\delta_{t h}\delta_{u v}):
@asym!(V3t3){_{p},_{t},_{u}}: @asym!(V3t3){_{f},_{g},_{h}}:
@asym!(V3t3){_{n},_{r},_{s}}: V3t:=@(V3t1)+@(V3t2)-@(V3t3):
@eliminate_kr!(V3t): V3:=V3^{m n r s p t u f g h}=@(V3t):
\end{cdbcont}
The 41st row to the 55th row define the symbols $V1^{abcde}$, $V2^{abcdefgh}$ and $V3^{mnrsptufgh}$ which correspond to the pure spinor correlations in Eqs.(\ref{PS1}-\ref{PS3}) respectively.
\begin{cdbcont}
COM:=(s_{1 1}->0,s_{2 2}->0,s_{3 3}->0,s_{4 4}->0,s_{5 5}->0,
\cdot{e^{1}}{k^{1}}->0,\cdot{e^{2}}{k^{2}}->0,\cdot{e^{3}}{k^{3}}->0,
\cdot{e^{4}}{k^{4}}->0,\cdot{e^{5}}{k^{5}}->0,\cdot{e^{1}}{k^{5}}->-
\cdot{e^{1}}{k^{2}}-\cdot{e^{1}}{k^{3}}-\cdot{e^{1}}{k^{4}},
\cdot{e^{2}}{k^{5}}->-\cdot{e^{2}}{k^{1}}-\cdot{e^{2}}{k^{3}}-
\cdot{e^{2}}{k^{4}},\cdot{e^{3}}{k^{5}}->-\cdot{e^{3}}{k^{1}}-
\cdot{e^{3}}{k^{2}}-\cdot{e^{3}}{k^{4}},\cdot{e^{4}}{k^{5}}->-
\cdot{e^{4}}{k^{1}}-\cdot{e^{4}}{k^{2}}-\cdot{e^{4}}{k^{3}},
\cdot{e^{5}}{k^{4}}->-\cdot{e^{5}}{k^{1}}-\cdot{e^{5}}{k^{2}}-
\cdot{e^{5}}{k^{3}}):
STU:=(s_{3 5}->s_{1 2}-s_{4 5}-s_{3 4},s_{1 3}->s_{4 5}-s_{1 2}-s_{2 3},
s_{2 5}->s_{3 4}-s_{1 2}-s_{1 5},s_{1 4}->s_{2 3}-s_{1 5}-s_{4 5},s_{2 4}->
s_{1 5}-s_{2 3}-s_{3 4}):
\end{cdbcont}
The 56th row to the 65th row give the input of replacement rule of on-shell identities $(k^{i})^{2}=0$, $k^{i}\cdot e^{i}=0$. Considering momentum conservation $\sum_{i}k^{i}=0$, the scalar products $e^{1}\cdot k^{5}$, $e^{2}\cdot k^{5}$, $e^{3}\cdot k^{5}$, $e^{4}\cdot k^{5}$ and $e^{5}\cdot k^{4}$ are expressed by other scalar products and not shown in the final result. The 66th row to the 68th row declare the replacement rule in Eq.(\ref{ss}).
\begin{cdbcont}
lamthe:=s_{1 2}**(-1)s_{4 5}**(-1) T_{1 2} V^{3} T_{4 5}:
@substitute!!(lamthe)(@(T12),@(L21),@(Vk)): @distribute!(lamthe):
@substitute!!(lamthe)(@(Vi),@(Aalpha),@(Am),@(Walpha)):
@distribute!(lamthe): @keep_weight!(lamthe){PS1}{3}:
@keep_weight!(lamthe){PS2}{5}:
\end{cdbcont}
The 69th row give the input of the first term in Eq.(\ref{tree5}), the result from other terms will be generated by cyclic permutations of the numbers $\{1,2,3,4,5\}$, which can decrease the execution time. The 70th row replaces the BRST building block $T_{12}$ and superfield $L_{12}$ with explicit expressions that consist
of the superfields $A_{\alpha}$, $A_{m}$ and $W_{\alpha}$. The 71st row replaces the superfields $A_{\alpha}$, $A_{m}$ and $W_{\alpha}$ with their $\theta$ expansions. The 72nd row and the 73rd row select the terms which have three $\lambda$'s and five $\theta$'s. 
\begin{cdbcont}
@combine!(lamthe):
@substitute!!(lamthe)(\indexbracket{(\lambda\Gamma^{a}\Gamma^{b c}\theta)}->
\indexbracket{(\lambda\Gamma^{a b c}\theta)}
+\delta_{a b}\indexbracket{(\lambda\Gamma^{c}\theta)}
-\delta_{a c}\indexbracket{(\lambda\Gamma^{b}\theta)}):
@distribute!(lamthe): @prodsort!(lamthe):
@substitute!!(lamthe)(\indexbracket{(\lambda\Gamma^{a}\theta)}->O_{1}^{a},
\indexbracket{(\lambda\Gamma^{a b c}\theta)}->O_{1}^{a b c},
\indexbracket{(\theta\Gamma^{a b c}\theta)}->O_{2}^{a b c}):
@substitute!!(lamthe)(
O_{1}^{a}O_{1}^{b}O_{1}^{c}O_{2}^{d e f}->V1^{a b c d e f},
O_{1}^{a}O_{1}^{b}O_{1}^{c d e}O_{2}^{f g h}->V2^{a b c d e f g h},
O_{1}^{m}O_{1}^{n r s}O_{1}^{p t u}O_{2}^{f g h}->V3^{m n r s p t u f g h}):
\end{cdbcont}
The 74th row contracts the symbols with Weyl spinor indices. At this step, the result has one non-standard term $(\lambda\Gamma^{a}\Gamma^{b c}\theta)$, and the 75th to the 78th row do the following replacement
\begin{eqnarray}
(\lambda\Gamma^{a}\Gamma^{b c}\theta)\rightarrow (\lambda\Gamma^{a b c}\theta)+\delta_{a b}(\lambda\Gamma^{c}\theta)-\delta_{a c}(\lambda \Gamma^{b} \theta)
\end{eqnarray}
The 79th row to 82nd row rewrite the amplitude with the notations in Eqs.(\ref{O1}-\ref{O3}). The 83rd row to the 86th row assemble the symbols $O^{a}_{1}$, $O^{abc}_{1}$ and $O^{abc}_{2}$ into the symbols $V1^{abcde}$, $V2^{abcdefgh}$ and $V3^{mnrsptufgh}$.
\begin{cdbcont}
@substitute!!(lamthe)(\partial_{a}{F^{?}_{b c}}
->k^{?}_{a} (k^{?}_{b}e^{?}_{c}-k^{?}_{c}e^{?}_{b})):
@substitute!!(lamthe)(@(V1),@(V2),@(V3)):
@canonicalise!(lamthe): @distribute!(lamthe):
@substitute!!(lamthe)(Z^{?}_{a}->e^{?}_{a},F^{?}_{a b}->(k^{?}_{a}e^{?}_{b}-
k^{?}_{b}e^{?}_{a})): @distribute!(lamthe):
@eliminate_kr!(lamthe): @prodsort!(lamthe):
@substitute!!(lamthe)(e^{?}_{a}k^{??}_{a}->\cdot{e^{?}}{k^{??}},
e^{?}_{a}e^{??}_{a}->\cdot{e^{?}}{e^{??}},k^{?}_{a}k^{??}_{a}->s_{? ??}):
@substitute!!(lamthe)(@(COM)): @substitute!!(lamthe)(@(STU)):
@distribute!(lamthe): @prodsort!(lamthe): @collect_factors!(lamthe):
lamtheini:=@(lamthe):
@substitute!!(lamtheini)(e^{1}->e^{r1},e^{2}->e^{r2},e^{3}->e^{r3},
e^{4}->e^{r4},e^{5}->e^{r5},k^{1}->k^{r1},k^{2}->k^{r2},k^{3}->k^{r3},
k^{4}->k^{r4},k^{5}->k^{r5},s_{1 2}->s_{r1 r2},s_{2 3}->s_{r2 r3},
s_{3 4}->s_{r3 r4},s_{4 5}->s_{r4 r5},s_{1 5}->s_{r1 r5}):
\end{cdbcont}
The 87th row and the 88th row replace the $\partial_{a} F_{bc}$ with explict expression $k_{a} (k_{b}e_{c}-k_{c}e_{b})$. The 89th row and the 90th row replace the symbols $V1^{abcde}$, $V2^{abcdefgh}$ and $V3^{mnrsptufgh}$ with the products of several Kronecker deltas in the right-hand of Eqs.(\ref{PS1}-\ref{PS3}). The 91st row and the 92nd row replace $Z_{a}$ and $F_{ab}$ with explicit expressions $e_{a}$ and $(k_{a}e_{b}-k_{b}e_{a})$. The replacement of $\partial_{a} F_{bc}$ and $F_{ab}$ are divided into two steps and this could decrease the execution time. The 93rd row contracts the Kronecker deltas with momenta $k^{i}_{a}$ and polarization vectors $e^{i}_{a}$ such that the Kronecker deltas are removed from the following result. The 94th row and the 95th row contracts the momenta $k^{i}_{a}$ and polarization vectors $e^{i}_{a}$. The 96th row applys the on-shell identities and momentum conservation to the result. In the 97th row, the Mandelstam variables with arbitrary powers are collected. At this step, the computation of first term in Eq.(\ref{tree5}) is completed. The next step is to calculate other terms in Eq.(\ref{tree5}) by cyclic permutations of the labels $\{1,2,3,4,5\}$. The 98th row to 102nd row replace the labels $\{1,2,3,4,5\}$ with the symbols $\{r1,r2,r3,r4,r5\}$. 
\begin{cdbcont}
lamthe2:=@(lamtheini):
@substitute!!(lamthe2)(^{r1}->^{2},^{r2}->^{3},^{r3}->^{4},^{r4}->^{5},
^{r5}->^{1},_{r1}->_{2},_{r2}->_{3},_{r3}->_{4},_{r4}->_{5},_{r5}->_{1}):
lamthe3:=@(lamtheini):
@substitute!!(lamthe3)(^{r1}->^{3},^{r2}->^{4},^{r3}->^{5},^{r4}->^{1},
^{r5}->^{2},_{r1}->_{3},_{r2}->_{4},_{r3}->_{5},_{r4}->_{1},_{r5}->_{2}):
lamthe4:=@(lamtheini):
@substitute!!(lamthe4)(^{r1}->^{4},^{r2}->^{5},^{r3}->^{1},^{r4}->^{2},
^{r5}->^{3},_{r1}->_{4},_{r2}->_{5},_{r3}->_{1},_{r4}->_{2},_{r5}->_{3}):
lamthe5:=@(lamtheini):
@substitute!!(lamthe5)(^{r1}->^{5},^{r2}->^{1},^{r3}->^{2},^{r4}->^{3},
^{r5}->^{4},_{r1}->_{5},_{r2}->_{1},_{r3}->_{2},_{r4}->_{3},_{r5}->_{4}):
lamthe15:=@(lamthe)+@(lamthe2)+@(lamthe3)+@(lamthe4)+@(lamthe5):
@substitute!!(lamthe15)(s_{5 1}->s_{1 5},s_{2 1}->s_{1 2},s_{3 2}->s_{2 3},
s_{4 3}->s_{3 4},s_{5 4}->s_{4 5},\cdot{e^{2}}{e^{1}}->\cdot{e^{1}}{e^{2}},
\cdot{e^{3}}{e^{1}}->\cdot{e^{1}}{e^{3}},\cdot{e^{4}}{e^{1}}->
\cdot{e^{1}}{e^{4}},\cdot{e^{5}}{e^{1}}->\cdot{e^{1}}{e^{5}},
\cdot{e^{3}}{e^{2}}->\cdot{e^{2}}{e^{3}},\cdot{e^{4}}{e^{2}}->
\cdot{e^{2}}{e^{4}},\cdot{e^{5}}{e^{2}}->\cdot{e^{2}}{e^{5}},
\cdot{e^{4}}{e^{3}}->\cdot{e^{3}}{e^{4}},\cdot{e^{5}}{e^{3}}->
\cdot{e^{3}}{e^{5}},\cdot{e^{5}}{e^{4}}->\cdot{e^{4}}{e^{5}}):
@substitute!!(lamthe15)(@(COM)):
@distribute!(lamthe15): @prodsort!(lamthe15);
\end{cdbcont}
The 103rd row to the 114th row replace the symbols $\{r1,r2,r3,r4,r5\}$ with $\{2,3,4,5,1\}$, $\{3,4,5,1,2\}$, $\{4,5,1,2,3\}$ and $\{5,1,2,3,4\}$, and the results correspond to the other four terms in Eq.(\ref{tree5}) respectively. In the 115th row, the result from the five terms in Eq.(\ref{tree5}) are summed up.  The 116th row to the 123rd row reorder the numbers in Mandelstam variables and scalar products of momenta and polarization vectors. The 124th row applys the on-shell identities to the result. Finally, the 125th row  generates the final result for 5-gluon superstring amplitude. Using the algorithm \verb|@timing|, one can check the execution time spent in the routine, and for our laptop the execution time is about 29 seconds. The final output is
\begin{flalign}
\hspace{3ex}
&e^{1}\cdot e^{3}e^{2}\cdot k^{1}e^{4}\cdot k^{3}e^{5}\cdot k^{2}s_{12}^{-1}s_{45}^{-1}-e^{1}\cdot e^{3}e^{2}\cdot e^{5}e^{4}\cdot k^{1}s_{45}^{-1}-e^{1}\cdot e^{3}e^{2}\cdot e^{5}e^{4}\cdot k^{2}s_{45}^{-1}&\nonumber\\
&
-e^{1}\cdot e^{3}e^{2}\cdot e^{5}e^{4}\cdot k^{3}s_{45}^{-1}-e^{1}\cdot e^{5}e^{2}\cdot k^{1}e^{3}\cdot k^{2}e^{4}\cdot k^{1}s_{12}^{-1}s_{45}^{-1}-e^{1}\cdot e^{5}e^{2}\cdot k^{1}
e^{3}\cdot k^{2}&\nonumber\\
&e^{4}\cdot k^{2}s_{12}^{-1}s_{45}^{-1}-e^{1}\cdot e^{5}e^{2}\cdot k^{1}e^{3}\cdot k^{2}e^{4}\cdot k^{3}s_{12}^{-1}s_{45}^{-1}+e^{1}\cdot e^{5}e^{2}\cdot e^{3}e^{4}\cdot k^{1}s_{45}^{-1}+\ldots,&
\end{flalign}
which is consistent with the result in \cite{PSS,PSSweb}.

\section{Conclusions}
\label{sec:5} 
In this article, we have displayed the complicated computations of the superstring scattering amplitudes in pure spinor formalism can be dealed with the algorithms in package Cadabra. Cadabra was designed as making problem solving resemble as close as possible the steps one would follow with pencil and paper. The input and output using Tex format make it easier for beginners. The performances, such as, the selecting of terms with $\langle\lambda^{3}\theta^{5}\rangle$, the calculations of pure spinor correlations in terms of Kronecker deltas and so on, are well done in Cadabra. However, there are also several inconveniences, such as, the complicated substitutions, the definition of tensor functions and the programming language. All of these are expected to be improved in the new version of Cadabra.

Recently, motivated by the computations in pure spinor formalism \cite{CM2010,CM2011,COS2013873,CO2014099,CM2008}, kinematic factors of the $n$-point string
amplitude at tree-level can always be written in terms of multiparticle vertex operators which are described by multiparticle superfields $K_{B}\in\{A^{B}_{\alpha}$, $A^{m}_{B}$,  $W^{\alpha}_{B}$, ${\cal F}^{B}_{mn}\}$, where $B=12...p$ \cite{CMOS2014,CMOS2015,LMS2016}. The $n$-point SYM tree-level amplitude with  multiparticle Berends-Giele currents is expressed in terms of the components $e_{B}$ and $\chi_{B}$ which depend on the single-particle gluon polarizations $e^{m}_{i}$ and the single-particle gluino polarizations $\chi^{m}_{i}$. All tree-level computations can be reduced to the $\langle\lambda^3 \theta^5 \rangle$ correlators from the 3-point function. These developments greatly simplify the extraction of components from pure spinor superspace expressions and lead to significant performance improvements for computer methods.

\section*{Acknowledgements}
We are indebted to Ricardo Medina and C. R. Mafra for valuable comments on a draft of this article. The work has been supported by the Natural Science Foundation of Hebei province with Grant No. A2016201069, the Science Research Foundation of China Three Gorges University with Grant No.KJ2015A007.

\end{document}